\documentclass[aps,prl,twocolumn]{revtex4}

\begin{document}
\title{Mesoscopic Luttinger Liquid Theory in an Aharonov-Bohm Ring}
\author{Mun Dae Kim$^{1,2}$, Sam Young Cho$^{1,2}$, Chul Koo Kim$^{1,2},$ and Kyun Nahm$^3$}
\affiliation{
        $^1$Institute of Physics and Applied Physics, Yonsei University,
          Seoul 120-749, Korea\\
        $^2$Center for Strongly Correlated Materials Research,
          Seoul National University, Seoul 151-742, Korea\\
        $^3$Department of Physics, Yonsei University, Wonju 220-710, Korea
        }
\date{\today}
\begin{abstract}
 A careful study on the mesoscopic persistent current  in a Luttinger liquid ring is carried out.
 It is shown that discreteness plays an important role in calculating the persistent current
  caused by the magnetic flux. At zero temperature, the current is shown to be independent of the
 interaction even when $g=(g_2-g_4)/2$ is not zero.  The current
 becomes enhanced at finite temperatures with respect to the non-interacting case,
 when the parameter $g$ is positive.
\end{abstract}
\maketitle

With rapid development of nano-fabrication technology much attention has been paid
to the  mesoscopic system which exhibits  strong quantum phase coherence. Many
efforts have been devoted to incorporate the interaction effect between electrons
because they are strongly correlated in low dimensions.
The Luttinger liquid(LL) theory\cite{Hal,Voit,Gogolin,Delft,Schulz}
is a standard method to investigate 1D interacting electron systems.
Therefore, it became a starting point
to study  the mesoscopic property such as the persistent current(PC)\cite{Loss}.

Here, we consider the zero mode current in a mesoscopic LL ring
without  the Aharonov-Bohm(AB) flux
and extend it to the persistent current produced by the AB flux.
In the continuum limit of  LL theory, the current is  calculated
through the continuity equation.
In this approach, the mean current carries a prefactor, $v_J$,
which is renormalized by interactions\cite{Hal}.
However, for a mesoscopic ring, the momentum level is discrete and,
thus, the continuum approach is not valid.
The mean current must be obtained from the zero mode.
We show that the zero mode current operator
does not depend on the interaction parameter.

In this study, we consider the
case when left-right symmetry is absent and show  the continuum
approximation\cite{Loss} is not appropriate for the PC in the Luttinger
liquid model, where discreteness plays a crucial role. In the
present calculation, we take discreteness into the calculation
explicitly and show that the calculation yields  new results
that, at zero temperature, the expectation value of the PC
does not depend on the LL interaction with $g\geq 0$.
At finite temperatures the interaction enhances the amplitude of PC
with respect to  that of the non-interaction case.

We, first, study the current in a mesoscopic LL ring without AB flux.
Here and after, we consider the spinless particles, because the spin
degrees do not give any qualitatively different results for the amplitude of the current.
The current can be defined through the continuity equation such that
$\frac{d}{dt}\rho(x)=i[H,\rho(x)]=\nabla j(x)$,
where $\rho(x)=\sum_pe^{ipx}\rho(p)\equiv-\frac1{\pi}\partial_x\varphi(x)$ is the bosonic density
operator in the LL.
Then $j(x)$ is  represented by
\begin{equation}
\label{current}
j(x)=i[H,\sum_p \frac1{ip}e^{ipx}\rho_p]+\tilde{c},
\end{equation}
where $\tilde{c}$ is an operator independent of $x$.
The mean current operator, $I$, can be defined by $I\equiv \frac1L \int^L_0dx'j(x')$.
In the limit when the level spacing $\Delta p\rightarrow 0$,
the mean current $I$ is given by
$I=\lim_{p\rightarrow 0}\frac1{pL}[H,\rho_p]$, which results that $I=v_J\frac{J}L$ with
the interaction dependent prefactor $v_J$.
However, for a mesoscopic wire, we cannot assume the continuum limit and, thus,
should sum over the discrete  $p$ values,
\begin{eqnarray}
I&=&\frac1L\int^L_0dx'\{i\sum_{p\neq 0}\frac1{ip}e^{ipx'}[H,\rho_p]
+i[H, \frac{N}L x']+\tilde{c}\}\nonumber\\
&=&\tilde{c},
\end{eqnarray}
since $[H,N]=0$.
Here, $I=\tilde{c}$ is the uniform zero mode current which can not be obtained through
the continuity equation, but only through  the definition in terms of total momentum of electrons.

When the total particle number $N_o$ is odd,
there is no current excitation at the ground state.
The number of particles at the right(left) branch, $N_{o+}(N_{o-})$, becomes
$N_{o+}=N_{o-}=(N_o-1)/2$, where we subtracted 1,
the number of particle at the zero momentum state.
For even $N_o$, $N_{o+}=(N_o-1+k_J)/2$ and $N_{o-}=(N_o-1-k_J)/2$.
If the right(left) branch has one more particle at the ground state, $k_J=1(-1)$.
The case when $N_o$ is odd, corresponds to $k_J=0$.
At finite temperatures, current excitation number
$N_+(N_-)$ is added to $N_{o+}(N_{o-})$.
For the non-interacting case, the zero mode current, $I$,
can be calculated in terms of total momentum,
$-\frac eL\sum_{n=-N_{o-}-N_-}^{N_{o+}+N_+}\frac{2\pi}Ln$,
and represented by the excitation numbers,
\begin{eqnarray}
\label{current}
I=-\frac{e}L\frac{N_o}2\frac{2\pi}L(J+k_J)=-\frac{ev_F}L(J+k_J).
\end{eqnarray}
Here, $J\equiv N_+ - N_-$ is the current excitation number and
$N\equiv N_+ + N_-$  the charge excitation number.
For an interacting case, this range of summation remains correct, if internal
interactions do not change the total momentum of the system.
The zero mode of the Hamiltonian exactly satisfies this condition.
The interaction effect is included in obtaining
the expectation value, $\langle I\rangle$.
However, the current operator should have the interaction independent
fermi velocity, $v_F$,  as a prefactor instead of $v_J$.
Here we define the fermi velocity such that $v_F\equiv\frac{N_o}2\frac{2\pi}L$,
since this corresponds to the product of the particle number at one branch
with the momentum discreteness.


In the LL theory,
the non-zero and the zero mode in the
Hamiltonian are decoupled and the zero mode is given by
$\frac{\pi}{2L}(v_F+\frac{g_4-g_2}{2\pi})(J+\frac{2\phi}{\phi_o})^2 +
\frac{\pi}{2L}(v_F+\frac{g_4+g_2}{2\pi})N^2.$
The contribution of the
flux in the kinetic part comes from changes in the momenta of
electrons in accordance to the twisted boundary condition and
changes continuously. But, in the interaction part, excitation
energy is determined by the current and charge excitation number
which are integers. In the continuum field theory, the particle
distribution is continuous and, thus, the interaction energy
can also be continuous. However, since we are now studying the mesoscopic
regime, where the particle discreteness is crucial. We show in the
following that a small level shift without charge or current excitation
does not cause any change in the interaction energy.

In order to account the level discreteness correctly,
the twisted boundary condition should be implemented
 from the beginning of the bosonization process. Also, a careful analysis on the discreteness
 of the particle distribution is required, since the parity effect comes from the discreteness
 of the particle numbers. When the fermion field is expanded,
 $\psi_r(x)\equiv\left(\frac{2\pi}L\right)^\frac12 \sum_{k=-\infty}^\infty e^{-ikx} c_{kr},$
 the twisted boundary condition gives a condition that
 $k=\frac{2\pi}L(n+\phi/\phi_o)$\cite{Ando}. Here, $n$ is an integer.
 With this modification on k, one can
 proceed the bosonization  process exactly same as when $\phi=0$. Considering the flux range,
 $-\frac{\pi}L<\frac{\phi}{\phi_o}\frac{2\pi}L<\frac{\pi}L$, we note that positive magnetic
 flux causes an upward(downward) shift of the right(left) branch electrons.
 Now, the zero mode contribution to the kinetic energy excitation
  can be  obtained from the prescription $H^o_k=\sum_r H^o_{kr}$ with
 $H^o_{kr}=\sum k:c^\dagger_{kr}c_{kr}:$, where :: signifies the normal ordering.

To obtain the zero mode bosonic form in $H^o_k$,
we consider the ladder operator which increases the number of
electrons above the Fermi level, such that
$U_r|N_r,N_{-r}\rangle=|N_r+1,N_{-r}\rangle$. The ladder operator can be constructed as
$\sum_k c_{kp}^\dagger \delta\left[rk-\left(k_F+\frac{2\pi}L
(N_r-\frac12)\right)\right]$\cite{Hal,Voit},
where $N_r$ represents the $N_r$th excitation state in the r branch.
Here, the energies are measured from the reference level, $v_F\frac{\pi}L$.
The twisted boundary condition shifts the momentum levels by
$\frac{2\phi}{\phi_o}\frac{\pi}L$
and the ladder  operator creates particles at the shifted levels such that,
\begin{eqnarray}
\label{ladder}
U_r=\sum_k c_{kp}^\dagger \delta\left[rk-\left(k_F+\frac{2\pi}L
(N_r-\frac12+r\frac{\phi}{\phi_o})\right)\right].
\end{eqnarray}

 The zero mode part of the kinetic energy excitation  at each branch is given by
 \begin{eqnarray}
 \label{kinzero}
 H^o_{kr}&=&\frac{2\pi v_F}L\sum_{n=1}^{N_r} (n-\frac12+r\frac{\phi}{\phi_o})
       =\frac{\pi v_F}L N_r^2 +r\frac{2\pi v_F}L\frac{\phi}{\phi_o}N_r. \nonumber\\
 &&
 \end{eqnarray}
 For the even parity, application of the magnetic flux removes the degeneracy of the uppermost
 particle occupation. Since positive $\phi$ shifts the level to the right side, the uppermost
 particle takes a level at the left branch. Thus, the energy cost for the excitation in the left branch
 become increased by the level width, $v_F \frac{2\pi}L$, for each excited particle.
 Considering this effect and Eq. (\ref{kinzero}), we obtain for the kinetic part
 \begin{eqnarray}
 H^o_k=\frac{\pi v_F}{2L}N^2+\frac{\pi v_F}{2L}\left((J+k_J)^2+2(J+k_J)\frac{2\phi}{\phi_o}\right),
 \end{eqnarray}
 where $k_J=-1(+1)$ for positive(negative) $\phi$ for the even parity and 0 for the odd parity.

The zero mode in the interacting part, $H^o_{int}$, is directly deducible from the Hamiltonian.
The wave vector shift due to the twisted boundary condition makes change in the bosonic
density operator as
$\rho(p)=\sum_k :c^\dagger_{k+\frac{\phi}{\phi_o}\frac{2\pi}L+p}
c_{k+\frac{\phi}{\phi_o}\frac{2\pi}L}:$.
We, thus, can write $H^o_{int}$ for the $p=0$ mode from the Hamiltonian,
\begin{eqnarray}
&&\frac{g_2}{2L}\sum_{r,k,k'}
:c^\dagger_{r,k+\frac{\phi}{\phi_o}\frac{2\pi}L}
 c_{r,k+\frac{\phi}{\phi_o}\frac{2\pi}L}:
 :c^\dagger_{-r,k'+\frac{\phi}{\phi_o}\frac{2\pi}L}
 c_{-r,k'+\frac{\phi}{\phi_o}\frac{2\pi}L}:\nonumber\\
&&+\frac{g_4}{2L}\sum_{r,k,k'}
:c^\dagger_{r,k+\frac{\phi}{\phi_o}\frac{2\pi}L}
c_{r,k+\frac{\phi}{\phi_o}\frac{2\pi}L}:
:c^\dagger_{r,k'+\frac{\phi}{\phi_o}\frac{2\pi}L}
c_{r,k'+\frac{\phi}{\phi_o}\frac{2\pi}L}:,\nonumber\\
&&
\end{eqnarray}
where the normal ordering subtracts the infinite ground state density of type-r fermions.
We observe that
$N_r\equiv\sum_{r,k} :c^\dagger_{r,k+\frac{\phi}{\phi_o}\frac{2\pi}L}
c_{r,k+\frac{\phi}{\phi_o}\frac{2\pi}L}:$
is an integer number of the excited fermions for any value of $\phi$.
Also, $N_r$ is  a good quantum number because $[H,N_r]=0$
and does not change continuously as implied in the continuum approximation.
 Here, we note that the interaction parameter $g_2(g_4)$ is introduced to
 describe the forward scattering between different(same) branch particles excited from  the fermi level.
 In order to perform the thermal average of the PC, we consider
 the energy levels of the excited states. Because the excitation energy in the interaction part
 is completely determined by the charge excitation number, $N$, and the current excitation number, $J$,
 the parity  effect does not appear.
 Therefore, we obtain for the zero mode of the interaction part,
\begin{eqnarray}
H^o_{int}=\frac1{2L}\left(\frac{g_4+g_2}2\right)N^2+\frac1{2L}\left(\frac{g_4-g_2}2\right)J^2.
\end{eqnarray}
Here, flux $\phi$ does not affect the interaction Hamiltonian.
The PC is known to be determined by the current excitation number $J$
and the AB flux $\phi$. Since the charge excitation degree is decoupled from the current part,
it is sufficient to consider  the  total current excitation contribution only.
\begin{eqnarray}
H^o_{J}=\frac{v_F\pi}{2L}\left((J+k_J)^2+2(J+k_J)\frac{2\phi}{\phi_o}\right)+\frac{g}{2L}J^2,
\end{eqnarray}
where $g=\frac{g_4-g_2}{2}$.

In the LL Hamiltonian, the zero modes are decoupled from the non-zero mode.
Here, we write down the zero mode of the partition function
\begin{eqnarray}
\label{partition}
&&Z_o = C\sum_m e^{-\beta\left[\frac{\pi v_F}{2L}\left((2m+k_J)^2
+2(2m+k_J)\frac{2\phi}{\phi_o}\right)+\frac{g}{2L}(2m)^2\right]},\nonumber\\
&&
\end{eqnarray}
where terms irrelevant in the current calculation are absorbed in
constant C. When there is no charge excitation,  the eigenvalue of operator, $J$,
becomes an even integer, $2m$.
It is necessary to carry out  the exact summation process  instead of
the integration process of the continuum approximation to incorporate the AB flux
effect  and the parity effect correctly.
The result is given by
\begin{eqnarray}
\label{Jacobi}
Z_0&=&C \theta_3\left(\frac{i\pi v_F\beta}{L}(k_J+\frac{2\phi}{\phi_o}),
e^{-\frac{2\pi\eta v\beta}{L}}\right)
e^{-\frac{2\pi v_F \beta}{L}k_J \frac{\phi}{\phi_o}},\nonumber\\
&&
\end{eqnarray}
where C is a constant which does not depend on the flux $\phi$ and
$\theta_3(v,q)=\sum_{n=-\infty}^\infty q^{n^2} e^{i2nv}$ is the Jacobi
theta function.

The current can be obtained  as before calculating the total momentum,
$-\frac eL\sum_{n=-N_{o-}-N_-}^{N_{o+}+N_+}
\frac{2\pi}L\left(n+\frac{\phi}{\phi_o}\right),$
and represented by the excitation numbers,
\begin{eqnarray}
\label{current}
I=-\frac{e v_F}L\left(J+k_J+\frac{2\phi}{\phi_o}\right),
\end{eqnarray}
where $v_F=\frac{N_o}2\frac{2\pi}L$.
Here we note that $I_o\equiv\frac{e v_F}L=\frac{2\pi v_F}{L\phi_o}$ in unit $\hbar=1$.
With  Eq. (\ref{partition}) and Eq. (\ref{current}), we readily obtain
the expectation value of PC given by
\begin{eqnarray}
\langle I\rangle=-\frac{\partial}{\partial\phi}\left[-\frac1{\beta}\ln{Z_o}\right]
-I_o\frac{2\phi}{\phi_o},
\end{eqnarray}
which has a final form
\begin{eqnarray}
\label{TempCurrent}
&&\langle I\rangle=-I_o\left[k_J +\frac{2\phi}{\phi_o}-2\sum_{n=1}^{\infty}(-1)^n
\frac{\sinh(k_J+\frac{2\phi}{\phi_o})\frac{nT^*}T}
{\sinh\frac{\eta v}{v_F}\frac{nT^*}T}\right],\nonumber\\
&&
\end{eqnarray}
where $T^*\equiv\frac{2\pi v_F}{k_B L}$ is the characteristic temperature. Here, we used
\begin{eqnarray}
\label{differ}
\frac{\partial_\alpha\theta_3(i\alpha,q)}{\theta_3(i\alpha,q)} =4\pi
i\sum_{n=1}^{\infty}(-1)^n\frac{q^n}{1-q^{2n}}\sinh2\pi n\alpha,
\end{eqnarray}
which is valid when $|\alpha| \leq -\frac1{2\pi}\ln q$\cite{Bateman}.

When the interaction preserves  the left-right symmetry,
$g_2=g_4$ i.e. $g=0$\cite{Hal}, we observe that there is no contribution
from the interaction  at all temperatures.
This relation, $g=0$, is not satisfied generally. Chiral edge currents in a
quantum  Hall annulus gives an example\cite{Kettemann}.
Symmetry breaking due to an AB flux penetrating  annulus induces
the PCs. The inner and the outer edge currents
consist the left and the right branch respectively in a conventional
LL. The interaction strength between the edges can
be represented by the  $g_2$ parameter. Because the interedge
interaction is weaker than the intraedge one, we observe that $g>0$.

In order to determine the PC when $g \neq 0$, it is necessary to
determine the possible values of $k_J$.
Using Eq. (\ref{Jacobi}) and  the valid range of Eq. (\ref{differ})
with $\alpha=\frac{\pi v_F\beta}{L}(k_J+\frac{2\phi}{\phi_o})$ and
$q=\exp(-\frac{2\pi\eta v\beta}L)$,
we find $k_J$=0 or $-1$ for $0<\phi/\phi_o<1/2$ and $k_J$=0 or 1 for
$-1/2<\phi/\phi_o<0$.
The constraint on the current number further reduces to
$k_J=0$ for the odd parity and $k_J=-1(+1)$ for positive(negative)
$\phi$ for the even parity. We see that this is consistent with the values
of $k_J$ which we obtained from the physical argument previously.

We examine the zero temperature limit of the PC for positive $g$ using Eq. (\ref{TempCurrent})
and obtain results which are same to those in non-interacting case\cite{Cheung}:
\begin{eqnarray}
\langle I\rangle&=&-\frac{4\pi v_F\phi}{L\phi_o^2},~~~~~~~~~~~~~~~~~~~~
\mbox{odd parity},\nonumber\\
\langle I\rangle&=&\frac{2\pi v_F }{L\phi_o}(1-\frac{2|\phi|}{\phi_o})\rm{sgn}\phi,~~~~~
 \mbox{even parity}.
\end{eqnarray}
 The above results clearly show that the PC is not renormalized by interaction
at zero temperature contrary to the results of previous calculations\cite{Loss}.
This result is consistent with microscopic derivations\cite{Groeling}.

For finite temperatures, the PC shows a
qualitatively different behavior from the non-interacting case as
shown in Fig. 1. We, first, consider the even parity case. It
is sufficient to consider the positive $\phi$ regime($k_J=-1$).
We can see from the partition function in Eq. (\ref{partition}) that,
in the non-interacting case at the ground state, $m=0$.
The first excited state, $m=+1$, has a lower energy than that of the state, $m=-1$.
This reduces the value of current because value of the current
has the opposite sign of the momentum as shown in Eq. (\ref{current}).
 The interacting term, however, does not favor any particular current direction
 because of the equal energy cost for $m=\pm 1$. Therefore, inclusion of the interaction term
 makes the current reduction smaller and, thus, the amplitude is enhanced compared to the
 non-interacting case(Fig. 2).
For the odd parity, the PC is just the translation of the even parity case by
$\phi_o/2$ as can be seen in Eq. (\ref{TempCurrent}) with $k_J$=0.
The reduction of amplitude is also smaller than
that of the non-interacting case.
We note that the translational symmetry
between the odd and the even parity case and the
magnetization direction due to the PC are consistent with the
Legget's theorem\cite{Legget}.

\begin{figure}[t]
  \vspace*{11cm}
  \includegraphics{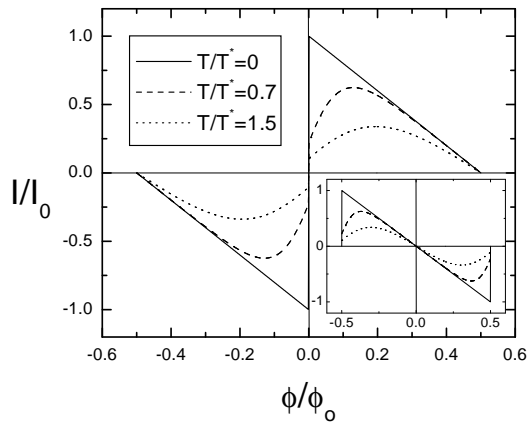}
  \vspace*{-4.5cm}
 \caption{PCs at several temperatures for the even parity, when $g=0.05$.
 The figure in the inset shows the odd parity case.}
 \label{fig1:model}
 \vspace*{1cm}
 \end{figure}

 \begin{figure}[]
  \vspace*{1.5cm}
 \includegraphics{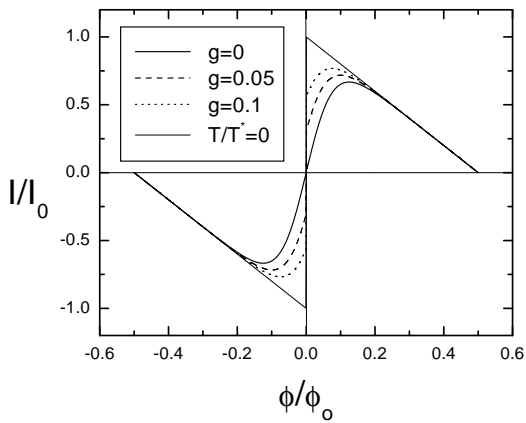}
  \vspace*{4.3cm}
 \caption{PCs for several interaction parameter $g$ values at $T/T^*=0.5$.
  The amplitudes of PCs are enhanced as g increases. The zero temperature current
  for an arbitrary $g$ is  a guide for eye. }
 \label{fig2:model}
 \end{figure}

Several groups have carried out
numerical calculations  for PCs in interacting systems with disorder\cite{Lanczos,Monte,DMRG}.
In those cases  interactions
preserve the left-right symmetry and, thus, they correspond to the case, $g=0$ in LL scheme.
Lanczos method\cite{Lanczos} showed that, for a clean ring, the amplitude of PC is independent
of the interaction strength away from half-filling. In the presence of disorder, however, the amplitude
become dependent on the interaction strength. Lanczos method for a dirty ring\cite{Lanczos} and Monte Carlo
simulation for a electron interacting with a diffuse environment\cite{Monte} demonstrated that the amplitude
is supressed, while it is enhanced in DMRG calculation for a disordered  ring\cite{DMRG}.

In summary, we have developed a LL theory in a mesoscopic ring.
In mesoscopic systems like AB ring, the level discreteness plays an important role.
We obtained the zero mode in the LL Hamiltonian
by exact summation instead of continuum integration.
At zero temperature, the PC
generally does not depend on the interaction, which is consistent with
microscopic derivations. At finite temperatures, amplitude of the PC is
enhanced with respect to the non-interacting case and its behavior confirms the
Legget's theorem.

This work was partly supported by the Korea Research Foundation(99-005-D00011)
and the Korea Science and Engineering Foundation through Center for Strongly
Correlated Materials Research(SNU).



\begin{thebibliography}{1}
\bibitem{Hal} F. D. M. Haldane, J. Phys. C {\bf 14}, 2585(1981).
\bibitem{Voit} J. Voit, Rep. Prog. Phys. {\bf 58}, 977(1995).
\bibitem{Gogolin} A. O. Gogolin, A. A. Nersesyan, and A. M. Tsvelik,
{\it Bosonization and Strongly Correlated Systems}(Cambridge University
Press, Cambridge, 1998).
\bibitem{Delft} J. von Delft and H. Schoeller, Ann. Phys.(Leipzig) {\bf 7}, 225(1998).
\bibitem{Schulz} H. J. Schulz, G. Guniberti, and P. Pieri,
 in {\it Field Theories for Low-Dimensional
 Condensed Matter Systems}, edited by G. Morandi {\it et al.}(Springer, 2000).
\bibitem{Loss} D. Loss, Phys. Rev. Lett. {\bf 69}, 343(1992);
 D. Schmeltzer, Phys. Rev. B {\bf 47}, 7591(1993);D. Schmeltzer and R. Berkovits,
Phys. Lett. A {\bf 253}, 341(1999);D. Schmeltzer, Phys. Rev. B {\bf 63}, 125332(2001).
\bibitem{Byers} N. Byers and C. N. Yang, Phys. Rev. Lett. {\bf 7}, 46(1961).
\bibitem{Cheung} H.-F. Cheung, Y. Gefen, E. K. Riedel and W.-H. Shih,
 Phys. Rev. B. {\bf 37}, 6050(1988);
 D. Loss and Goldbart, {\it ibid}. {\bf 43}, 13762(1991).
\bibitem{Mailly} L. P. L{\' e}vy, G. Dolan, J. Dunsmuir, and H. Bouchiat, Phys. Rev. Lett. {\bf 64},
 2074(1990); V. Chandrasekhar, R. A. Webb, M. J. Brady, M. B. Ketchen,
 W. J. Gallagher, and A. Kleinsasser, {\it ibid}. {\bf 67}, 3578(1991);
 D. Mailly, C. Chapelier, and A. Benoit, {\it ibid}. {\bf 70}, 2020(1993).
\bibitem{Ando}See, for instance, H. Fukuyama, in {\it Mesoscopic Physics and Electronics},
edited by T. Ando {\it et al.}(Springer, Berlin, 1998).
\bibitem{Bateman} H. Bateman and A. Erderly, {\it Higher Transcendental
 Functions}(McGraw Hill, 1954).
\bibitem{Kettemann} D. J. Thouless and Y. Gefen, Phys. Rev. Lett. {\bf 66}, 806(1991);
Y. Gefen and D. J. Thouless, Phys. Rev. B {\bf 47}, 10423(1993);
S. Kettemann, {\it ibid}. {\bf 55}, 2512(1997).
\bibitem{Groeling} A. M{\"u}ller-Groeling, H. A. Weidenm{\"u}ller
 and C. H. Lewenkopf, Europhys. Lett. {\bf 22}, 193(1993);
 A.M{\"u}ller-Groeling and H. A. Weidenm{\"u}ller, Phys. Rev. B. {\bf 49}, 4752(1994).
\bibitem{Legget} A. J. Legget, in {\it Granular Nanoelectronics}, edited by
 D. K. Ferry, J. R. Barker, and C. Jacoboni, NATO ASI Ser. B {\bf 251},
 279(Plenum, New York, 1991).
\bibitem{Lanczos} G. Bouzerar, D. Poilblanc, and G. Montambaux, Phys. Rev. B {\bf 49}, 8258(1994).
\bibitem{Monte} D. S. Golubev, C. P. Herrero, and A. D. Zaikin, cond-mat/0205549.
\bibitem{DMRG} P. Schmitteckert, R. A. Jalabert, D. Weinmann, and J.-L. Pichard, Phys. Rev. Lett.
{\bf 81}, 2308(1998).
\end{thebibliography}
\end{document}